# Reflection and refraction of an electron spin at the junction between two quasi-two-dimensional semiconductor regions with and without spin-orbit interaction

**Supriyo Bandyopadhyay[1], Marc Cahay[2] and Jonathan Ludwick[2]**

[1] Department of Electrical and Computer Engineering, Virginia Commonwealth University, Richmond, VA 23284, USA

[2] Department of Electrical Engineering and Computer Science, University of Cincinnati, Cincinnati, OH 45221, USA

E-mail: sbandy@vcu.edu

**Abstract**

We derive the reflection and refraction laws for an electron spin incident from a quasi-two-dimensional medium with no spin-orbit interaction on another with both Rashba and Dresselhaus spin-orbit interaction using only energy conservation. We obtain the well-known result that for an incident angle, there can be generally two different refraction angles for refraction into the two spin eigenstates in the refraction medium, resulting in two different 'spin refractive indices' and two critical angles for total internal reflection. We derive expressions for the spin refractive indices, which are not constant for a given medium but depend on the incident electron's energy. If the effective mass of an electron in the refraction medium is larger than that in the incidence medium, then we show that for some incident electron energies and potential barrier at the interface, the spin refractive index of the incidence medium can lie between the two spin refractive indices of the refraction medium, resulting in only one critical angle. In that case, if the incident angle exceeds that critical angle, then refraction can occur into only one spin eigenstate in the refraction medium. If the system is engineered to make this happen, then it will be possible to obtain a very high degree of spin-polarized injection into the refraction medium. The amplitudes of reflection of the incident spin into its own spin eigenstate and the orthogonal spin eigenstate (due to spin flip at the interface), as well as the refraction amplitudes into the two spin eigenstates in the refraction medium are derived for an incident electron (with arbitrary spin polarization and incident energy) as a function of the angle of incidence.

## 1. Introduction

Spin-polarized injection of electrons from one quasi-two-dimensional medium into another (the latter possessing strong spin-orbit interaction) is at the heart of a number of spin field effect transistor ideas based on gate-tunable spin-orbit interaction [1, 2]. Normally, the injecting medium will a ferromagnet or half-metal with a high degree of spin-polarization. This strategy of spin-injection is unfortunately plagued by the resistance mismatch problem [3] and, even otherwise, there are very few ferromagnets or half-metals with sufficient spin polarization at room temperature to act as an efficient injector. Although the resistance mismatch problem can be ameliorated by interposing a tunnel barrier between the two media [4], this approach increases the source resistance and requires additional processing steps in spin field effect transistors, which are both undesirable. There are other strategies for spin-polarized injection (see, for example, [5]), but they all present their own challenges.

Recently, another modality of spin polarized injection has gained attention [6-9]. When an electron is incident from a quasi-two-dimensional (2-D) medium with no spin-orbit interaction on another quasi-2-D medium with Rashba and/or Dresselhaus spin-orbit interaction, there can be two angles of refraction into the second medium, one for each spin eigenstate (in the latter medium). This birefringence causes two spatially separated transmitted beams, one for each spin state. It is reminiscent of the Stern-Gerlach experiment where mutually antiparallel spins were spatially separated. This also means that the refraction medium has two different "spin refractive indices" – $n^+$ and $n^-$ – one for each spin eigenstate. The spatial separation allows one to resolve the two spins and collect them separately with collectors placed at two different locations. However, this does not implement "spin-polarized injection" into the medium with strong spin-orbit interaction and hence is not helpful for spin field effect transistors [1, 2].

To achieve spin-polarized injection into the refraction medium, one can further tailor this process to allow the incident electron to transmit into only one of the spin eigenstates in the refraction medium and not the other. This will happen if the two spins have two different critical angles of total internal reflection (which they must since $n^+ \neq n^-$) and the angle of incidence lies between these two values. An even better situation will arise if there is no real solution for one of the two critical angles, meaning that there is only one critical angle. This will require the condition $n_{II}^- \geq n_I \geq n_{II}^+$ where the refraction medium is labeled with subscript II and the incident medium with subscript I. In this case, the angle of incidence has to exceed only the one critical angle and that will ensure that the electron can transmit into only one of the two mutually orthogonal spin eigenstates in the refraction medium. This phenomenon can be leveraged to achieve a very high degree of spin-polarized injection. Here, we establish the conditions for this to happen by invoking only energy conservation in the reflection/refraction process.

## 2. Theory

Consider the interface between two quasi-two-dimensional (2-D) regions I and II, as shown in Fig. 1, where region II alone has Rashba and Dresselhaus spin-orbit interactions. An electron is incident from region I on region II. Without loss of generality, we will consider that only the lowest subband is occupied by electrons in either quasi-2-D region.

The Hamiltonian for the composite system can be written as

$$H = p_x \frac{1}{2m(x)} p_x + \frac{p_z^2}{2m(x)} + V(x) \\ -(\eta/\hbar)[p_x\sigma_z - p_z\sigma_x] + (\nu/\hbar)[p_x\sigma_z - p_z\sigma_z] \qquad (1)$$

where $\eta$ is the strength of the Rashba interaction and $\nu$ is the strength of the Dresselhaus interaction in region II (they have non-zero values only in region II). Since the Hamiltonian is invariant in the z-coordinate, the wavevector component $k_z$ is a good quantum number.

Note that we ignore the cubic Dresselhaus term in Equation (1). This term can make a difference in some cases [10], but it is usually not important in quasi-2D systems. In a quasi-2-D system, the cubic term becomes comparable to the linear term when $k_x k_z \geq (\pi/d)^2$, where $k_x$ and $k_z$ are the in-plane wavevector components and $d$ is the thickness of the 2-D layer. In 2-D layers grown by techniques such as molecular beam epitaxy, the layer can be a monolayer of thickness ~0.5 nm, in which case, the cubic term will become important for wavevectors exceeding $6 \times 10^9$ m$^{-1}$. The electron energies corresponding to such wavevectors will be ~ 3 eV assuming an effective mass that is one-half of the free electron mass. A Fermi wavevector of $6 \times 10^9$ m$^{-1}$ would correspond to a sheet carrier concentration of ~ $6 \times 10^{18}$ m$^{-2}$ which is unachievable in most 2-D systems. Hence, the cubic term is not likely to be important in our case.

Another effect we ignore is interfacial spin-orbit interaction. If there is a step discontinuity in the conduction band at the interface of the two materials, it will give rise to a delta-function electric field at the interface which will cause an additional delta-function Rashba interaction at the interface. That interaction can cause spin mixing, making the eigenspinors in region II spatially variant. Such effects cannot be included within an analytical analysis, and is hence ignored here. This will limit the validity of our analysis to systems where the step discontinuity is relatively small.

Let us assume that in region I, the effective mass is $m_I$ and in region II, it is $m_{II}$. We will also assume a step discontinuity in the potential at the interface, so $V(x) = V$ in region I and $V(x) = 0$ in region II. If we ignore band bending due to spaces charges, then $V$ is the conduction band offset at the interface.

Neglecting spin mixing effects (which could make the eigenspinors in region II vary with *x* and *z* coordinates), the energy dispersion relations of the lowest spin-split subband in region II are given by [11]

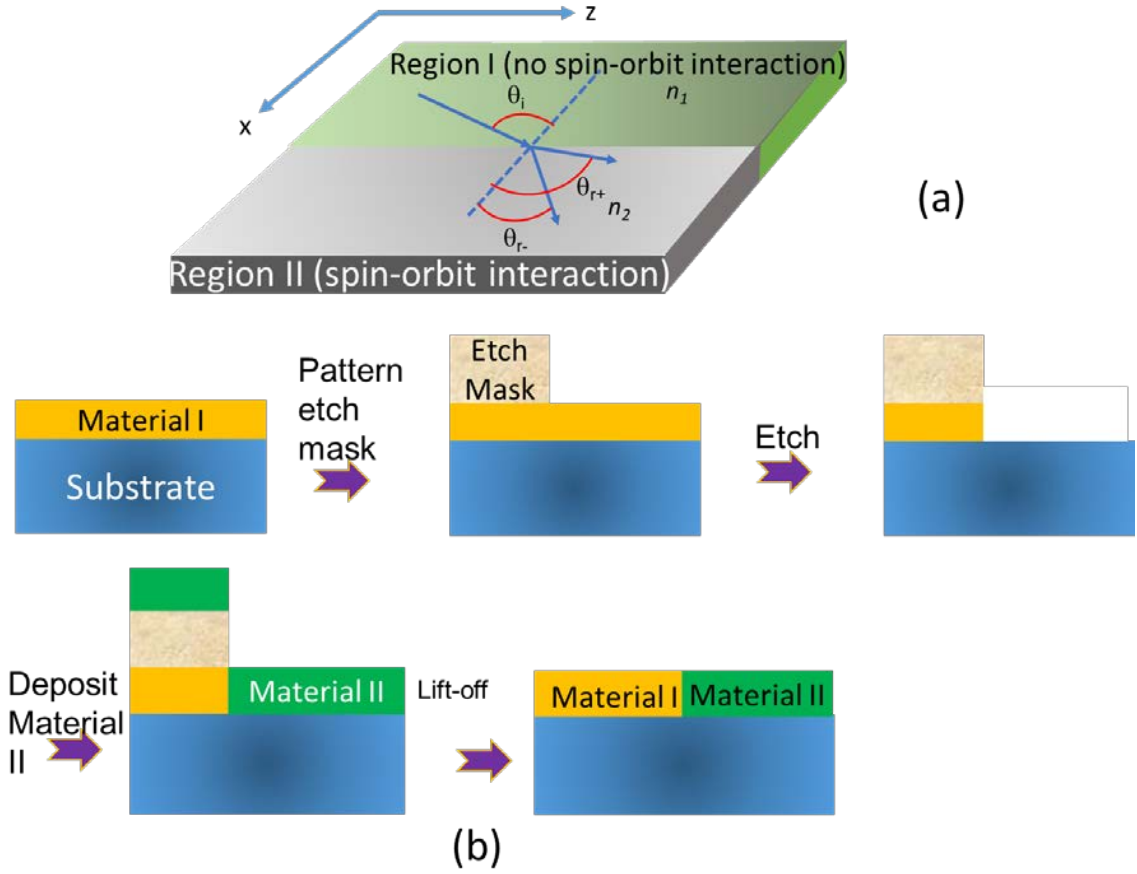

**Fig. 1:** (a) Interface between two two-dimensional regions, one of which has no spin-orbit interaction and the other has Rashba and Dresselhaus spin-orbit interactions. (b) Fabrication steps.

$$E_{\pm}^{II} = \frac{\hbar^2 \left(k_x^2 + k_z^2\right)}{2m_{II}} \pm \sqrt{\left(\eta k_x - \nu k_z\right)^2 + \left(\eta k_z - \nu k_x\right)^2} , \qquad (2)$$

where $k_x$ and $k_z$ are the wavevector components of an electron in region II.

Let us call the wavevector components in region I $\left(k_x', k_z'\right)$. The "spin refractive indices" of the two regions will obey the relation

$$\frac{k_I}{k_{II}} = \frac{\sqrt{k_x'^2 + k_z'^2}}{\sqrt{k_x^2 + k_z^2}} = \frac{n_I}{n_{II}} \neq \frac{v_I}{v_{II}} ,$$

$$E^I = E_{\pm}^{II} = E \qquad (3)$$

where $k_I(v_I)$ and $k_{II}(v_{II})$ are the magnitudes of the electron wave vectors(velocities) for an arbitrary electron energy $E$ in the two regions. The last inequality shows that the refractive index ratio does not exactly obey the relation of conventional optics, because in the presence of spin-orbit interaction, electron velocity is not proportional to the wavevector.

It is easy to see from the dispersion relations in Equation (2) that if only Rashba or only Dresselhaus interaction is present, then both spin components travel with the same speed in the refraction medium, albeit in different directions [6].

Using Equation (3), we will now define a quantity $k_0$ as $k_0 = \frac{\sqrt{k_x'^2 + k_z'^2}}{n_I} = \frac{\sqrt{k_x^2 + k_z^2}}{n_{II}}$.

The angles of incidence $\theta_i$ and refraction $\theta_r$ are defined in Fig. 1. We note from this figure that $k_x' = n_I k_0 \cos\theta_i, k_z' = n_I k_0 \sin\theta_i, k_x = n_{II} k_0 \cos\theta_r, k_z = n_{II} k_0 \sin\theta_r$. Since $k_z$ is a good quantum number, it is conserved across the interface, and hence $k_z' = k_z$, which immediately yields the equivalent of Snell's law from the last set of relations

$$\frac{\sin\theta_i}{\sin\theta_r} = \frac{n_{II}}{n_I} \quad . \quad (4)$$

It is also easy to see from the conservation of the wavevector's $z$-component that the angle of reflection is always equal to the angle of incidence in region I (which has no spin-obit interaction). This is identical to the situation in conventional optics.

The energy dispersion relation in region I is

$$E^I = \frac{\hbar^2 \left(k_x'^2 + k_z'^2\right)}{2m_I} + V = \frac{\hbar^2 n_I^2 k_0^2}{2m_I} + V . \quad (5)$$

Since energy is conserved in the process of reflection/refraction (which are elastic events), we must have $E^I = E_\pm^{II} = E$ which yields

$$\frac{\hbar^2 \left(k_x'^2 + k_z'^2\right)}{2m_I} + V = E$$
$$= \frac{\hbar^2 \left(k_x^2 + k_z^2\right)}{2m_{II}} \pm \sqrt{(\eta k_x - \nu k_z)^2 + (\eta k_z - \nu k_x)^2} \quad (6)$$

This equation immediately shows that there will be generally two different solutions for $k_x$ for the two spin eigenstates at any incident energy $E$ and hence, according to Equations (3) and (5), there will be two different refractive index $n_{II}^\pm$ for refracting into the two spin eigenstates in region II. Snell's law will then ensure that for a fixed angle of incidence, there will be two different refraction angles $\theta_r^\pm$ which will obey the relation

$$\frac{\hbar^2 n_I^2 k_0^2}{2m_I} + V = \frac{\hbar^2 \left(n_{II}^\pm\right)^2 k_0^2}{2m_{II}}$$
$$\pm n_{II}^\pm k_0 \sqrt{\left(\eta \sin\theta_r^\pm - \nu \cos\theta_r^\pm\right)^2 + \left(\eta \cos\theta_r^\pm - \nu \sin\theta_r^\pm\right)^2}, \quad (7)$$
$$= \frac{\hbar^2 \left(n_{II}^\pm\right)^2 k_0^2}{2m_{II}}$$
$$\pm n_{II}^\pm k_0 \sqrt{\eta^2 + \nu^2} \sqrt{\sin^2\left(\theta_r^\pm - \zeta\right) + \cos^2\left(\theta_r^\pm + \zeta\right)}$$

where $\zeta = \arctan(v/\eta)$.

The last equation follows from Equations (3), (4) and (6). The two signs in the equation above correspond to the two spin eigenstates in region II. Using Equation (4), we can recast Equation (7) as

$$1 + \frac{2m_1 V}{\hbar^2 n_I^2 k_0^2} = \frac{m_I}{m_{II}} \frac{\sin^2 \theta_i}{\sin^2 \theta_r^\pm} \pm \frac{\sin \theta_i}{\sin \theta_r^\pm} \frac{2m_I}{\hbar^2 n_I k_0} \sqrt{\eta^2 + v^2 - 2\eta v \sin(2\theta_r^\pm)} \qquad (8)$$

Equation (8) is easily solved for the refraction angles (for a given incident angle) when either Rashba or Dresselhaus spin-orbit interaction (but not both) is present. When only Rashba interaction is present, the solution is

$$\sin \theta_r^\pm = \frac{\sin \theta_i}{\mp \frac{m_{II} \eta}{\hbar^2 n_I k_0} + \sqrt{\left(\frac{m_{II} \eta}{\hbar^2 n_I k_0}\right)^2 + \frac{m_{II}}{m_I}\left(1 + \frac{2m_1 V}{\hbar^2 n_I^2 k_0^2}\right)}}$$

$$= \frac{\sin \theta_i}{\frac{m_{II} \eta}{\hbar \sqrt{2m_I(E-V)}}\left(\sqrt{1 + \frac{2E\hbar^2}{m_{II} \eta^2}} \mp 1\right)} \qquad (9)$$

This is the law of refraction when only Rashba interaction is present. Similarly, when only Dresselhaus interaction is present, Equation (9) will be the solution with $\eta$ replaced by $v$. In Equation (9), we rejected the negative solution for the refraction angle as extraneous since "negative refraction", which is well known in optics, is not relevant here. This equation shows that there are two refraction angles for a given angle of incidence $\theta_i$. It is also easy to show from Equation (9) that

$$\sin \theta_r^+ - \sin \theta_r^- = \sin \theta_i \frac{\eta \sqrt{2m_I(E-V)}}{\hbar E} \quad \text{(only Rashba present)}$$

$$= \sin \theta_i \frac{v \sqrt{2m_I(E-V)}}{\hbar E} \quad \text{(only Dresselhaus present)} \qquad (10)$$

which immediately shows that $\theta_r^+ \geq \theta_r^-$. This then also means that $n_{II}^- \geq n_{II}^+$.

If we have only Rashba interaction, then it is easy to show from Equations (4) and (9) that

$$\frac{n_{II}^\pm}{n_I} = \frac{\eta m_{II}}{\hbar} \frac{\sqrt{1 + \frac{2E\hbar^2}{m_{II}\eta^2}} \mp 1}{\hbar \sqrt{2m_I(E-V)}} \quad \text{(only Rashba present)}$$

$$= \frac{v m_{II}}{\hbar} \frac{\sqrt{1 + \frac{2E\hbar^2}{m_{II}v^2}} \mp 1}{\hbar \sqrt{2m_I(E-V)}} \quad \text{(only Dresselhaus present)} \qquad (11)$$

which shows that the ratio of the spin refractive indices in the two media depend on the effective masses, the incident energy of the electron, the potential barrier height at the interface and the strength of the spin-orbit interaction. It is interesting to note that when only Rashba or only Dresselhaus interaction is present, $\frac{n_{II}^+}{n_I} \times \frac{n_{II}^-}{n_I} = \frac{m_{II}}{m_I} \frac{E}{E-V}$, which is independent of the spin-orbit interaction strength.

When both Rashba and Dresselhaus interactions are present, the spin refractive index ratio $n_{II}^{\pm}/n_I$ must be found by solving Equations (4) and (8). Curiously, in this case, the relative spin refractive index of region II, i.e. $n_{II}^{\pm}/n_I$, will depend on the angle of incidence. This has no analogy in conventional optics.

To find the critical angle(s) for total internal reflection associated with the first spin eigenstate in the refraction medium (labeled by the + sign), we set $\theta_r^+ = 90^0$ in Equation (8) and solve for $\theta_i$. This yields (we always take the positive value of the radical)

$$1 + \frac{2m_I V}{\hbar^2 n_I^2 k_0^2} = \frac{m_{II}}{m_I} \sin^2 \theta_c^+ + \frac{2m_I}{\hbar^2 n_I k_0} \sqrt{\eta^2 + v^2} \sin \theta_c^+, \tag{12}$$

which yields two different solutions for the critical angle $\theta_c^+$. The first is

$$\theta_{c1} = \arcsin\left(-\frac{m_{II}}{\hbar^2 n_I k_0}\sqrt{\eta^2+v^2} + \sqrt{\left(\frac{m_{II}}{\hbar^2 n_I k_0}\right)^2 (\eta^2+v^2) + \frac{m_{II}}{m_I} + \frac{2m_{II}V}{\hbar^2 n_I^2 k_0^2}}\right),$$

$$= \arcsin\left(\frac{m_{II}\sqrt{\eta^2+v^2}}{\hbar} \frac{\sqrt{1+\frac{2E\hbar^2}{m_{II}(\eta^2+v^2)}} - 1}{\sqrt{2m_I(E-V)}}\right)$$

which reduces to the familiar expression (see Equation (11))

$$\theta_{c1} = \arcsin\left(\frac{n_{II}^+}{n_I}\right), \tag{13}$$

when only Rashba or only Dresselhaus interaction is present.

The other solution for the critical angle $\theta_c^+$ is

$$\theta_{c2} = \arcsin\left(-\frac{m_{II}}{\hbar^2 n_I k_0}\sqrt{\eta^2+v^2} - \sqrt{\left(\frac{m_{II}}{\hbar^2 n_I k_0}\right)^2 (\eta^2+v^2) + \frac{m_{II}}{m_I} + \frac{2m_{II}V}{\hbar^2 n_I^2 k_0^2}}\right) \tag{14}$$

Since the angle $\theta_{c2}$ has a negative value, it is rejected as an extraneous solution. The angle $\theta_{c1}$ is positive and has a *real value* as long as

$$\frac{m_{II}\sqrt{\eta^2+v^2}}{\hbar} \frac{\sqrt{1+\frac{2E\hbar^2}{m_{II}(\eta^2+v^2)}} - 1}{\sqrt{2m_I(E-V)}} \leq 1$$

$$\Rightarrow \tag{15}$$

$$\frac{m_{II}}{m_I}\left(1 + \frac{V}{E-V} - \frac{\sqrt{\eta^2+v^2}}{\hbar}\sqrt{\frac{2m_I}{E-V}}\right) \leq 1$$

When taken with Equation (11), which is valid when only one type of spin-orbit interaction is present, the last equation implies that a critical angle exists only when $n_{II}^+ > n_I$, which is of course analogous to the situation in optics. Thus, a critical angle exists for an electron that would refract into the first spin eigenstate in the refraction medium if Equation (15) is satisfied. *If the incident angle is equal to or greater than this critical angle, then no transmission/refraction* *into the first spin eigenstate in region II can take place* which would then allow 100% spin polarized injection into region II. This is the basis of spin-polarized injection.

For the second spin eigenstate (labeled by the – sign), we again set $\theta_r^- = 90^0$ in Equation (8) and solve for $\theta_t$ to obtain the critical angle(s) for total internal reflection for a spin that would refract into the second spin eigenstate in region II. This yields (we always take the positive value of the radical)

$$1 + \frac{2m_I V}{\hbar^2 n_I^2 k_0^2} = \frac{m_I}{m_{II}} \sin^2 \theta_c^- - \frac{2m_I}{\hbar^2 n_I k_0} \sqrt{\eta^2 + \nu^2} \sin \theta_c^-, \tag{16}$$

which provides two different values for the critical angle $\theta_c^-$. The first is

$$\theta_{c3} = \arcsin\left( \frac{m_{II}}{\hbar^2 n_I k_0} \sqrt{\eta^2 + \nu^2} + \sqrt{\left(\frac{m_{II}}{\hbar^2 n_I k_0}\right)^2 (\eta^2 + \nu^2) + \frac{m_{II}}{m_I} + \frac{2m_{II} V}{\hbar^2 n_I^2 k_0^2}} \right)$$

$$= \arcsin\left( \frac{m_{II} \sqrt{\eta^2 + \nu^2}}{\hbar} \cdot \frac{\sqrt{1 + \frac{2E\hbar^2}{m_{II}(\eta^2 + \nu^2)}} + 1}{\sqrt{2m_I (E - V)}} \right)$$

which, of course, reduces to the familiar expression (see Equation (11))

$$\theta_{c3} = \arcsin\left( \frac{n_{II}^-}{n_I} \right), \tag{17}$$

when only Rashba or only Dresselhaus interaction is present.

The other solution for the critical angle $\theta_c^-$ is

$$\theta_{c4} = \arcsin\left( \frac{m_{II}}{\hbar^2 n_I k_0} \sqrt{\eta^2 + \nu^2} - \sqrt{\left(\frac{m_{II}}{\hbar^2 n_I k_0}\right)^2 (\eta^2 + \nu^2) + \frac{m_{II}}{m_I} + \frac{2m_{II} V}{\hbar^2 n_I^2 k_0^2}} \right) \tag{18}$$

Since the angle $\theta_{c4}$ has a negative value, it is rejected as an extraneous solution. The angle $\theta_{c3}$ is positive and has a *real value* as long as

$$\frac{m_{II}}{m_I} \left( 1 + \frac{V}{E - V} + \frac{\sqrt{\eta^2 + \nu^2}}{\hbar} \sqrt{\frac{2m_I}{E - V}} \right) \leq 1 \quad . \tag{19}$$

When taken with Equation (11), which is valid when only one type of spin-orbit interaction is present, the last equation implies that a critical angle exists only when $n_{II}^- > n_I$, which is again analogous to the situation in conventional optics. Thus, a critical angle exists for an electron that would refract into the second spin eigenstate in the refraction medium if Equation (19) is satisfied. *If the incident angle is equal to or greater than this critical angle, then no transmission/refraction into the second spin eigenstate in region II can take place.*

From Equations (13) and (17), we see that $\theta_{c3} > \theta_{c1}$ since we have shown that $n_{II}^- \geq n_{II}^+$. If *both* equations (15) and (19) are satisfied, i.e. both critical angles exist, then: (i) refraction into both spin eigenstates in region II will occur if $\theta_i < \theta_{c1}$, (ii) refraction into only the second spin eigenstate will occur if $\theta_{c3} > \theta_i > \theta_{c1}$, and (iii) no refraction will occur (i.e. there will be complete total internal reflection) if $\theta_i > \theta_{c3}$.

We will now examine two cases, corresponding to two different band alignments at the interface of regions I and II, to determine if both critical angles can exist, i.e. if both Equations (15) and (19) – or only one of them – can be satisfied.

**Case I: The conduction band edge in region I is above that in region II, or is in alignment with it, (i.e. $V \geq 0$) and $m_{II} \geq m_I$ (electron's effective mass in the refraction medium is heavier)**

In this case, Equation (19) can never be satisfied and hence the angle $\theta_{c3}$ will *not* have a real solution. In other words, if the effective mass of an electron in the refraction medium exceeds or equals that in the medium of incidence, then, for this band alignment, an electron incident from region I will always be able to transmit into the second spin eigenstate in region II, regardless of the angle of incidence, since it can never suffer total internal reflection back into region I.

By enforcing this condition on the effective masses for this band alignment, we can *completely eliminate* the possibility of total internal reflection for an electron attempting to couple into *one* of the two spin eigenstates in the refraction medium. Total internal reflection, however, will be possible for the other spin eigenstate (i.e. $\theta_{c1}$ can have real solutions), if Equation (15) is satisfied. By ensuring that Equation (15) is satisfied, we will be ensuring that transmission into only one spin eigenstate in the refraction medium is possible, while transmission into the other is blocked, as long as the angle of incidence exceeds $\theta_{c1}$. *This will allow for 100% spin polarized injection (or transmission) into the refraction medium if the angle of incidence $\theta_i > \theta_{c1}$.*

For the effective mass relationship and the band alignment we are considering here, Equation (15) can be satisfied only for certain electron energies $E$. In order to identify these electron energies, we define a quantity $X = 1/\sqrt{E-V}$, and then write Equation (15) as

$$f_1 = VX^2 - \frac{\sqrt{2m_I(\eta^2 + \nu^2)}}{\hbar}X + \left(1 - \frac{m_I}{m_{II}}\right) \leq 0. \qquad (20)$$

The function $f_1$ reaches its maximum value of $+\infty$ when $X = \infty$ and its minimum value of $-\frac{(\eta^2 + \nu^2)m_I}{2\hbar^2 V} + \left(1 - \frac{m_I}{m_{II}}\right)$ when $X = \frac{\sqrt{m_I(\eta^2 + \nu^2)}}{\sqrt{2}\hbar V}$ corresponding to the electron energy $E$ in region II equal to $(2\hbar^2 V^2)/(m_I[\eta^2 + \nu^2]) + V$. The function's maximum value obviously does not satisfy Equation (20), and its minimum value has to be zero or negative, according to Equation (20), in order that $\theta_{c1}$ can have a real value ever. The last condition translates to

$$\frac{\eta^2 + \nu^2}{2\hbar^2 V} \geq \frac{1}{m_I} - \frac{1}{m_{II}} \qquad (21)$$

Equation (21) is the condition for total internal reflection to be possible for an electron attempting to refract into the first spin eigenstate in the refraction medium. It becomes increasingly harder to satisfy this condition as the potential $V$ increases. Hence a smaller potential step is conducive to the critical angle having a real value. This equation also becomes harder to satisfy as the difference between the inverse of the effective masses increases. Weaker spin-orbit interaction strengths also make this equation harder to satisfy.

As long as Equation (21) is satisfied, $\theta_{c1}$ can have a real value for *some electron energy E* while $\theta_{c3}$ cannot have a real value. In other words, an electron attempting to refract into one spin state in region II can suffer total internal reflection if the incident angle exceeds $\theta_{c1}$ but never suffer total internal reflection when attempting to refract into the other spin state. This is actually equivalent to the inequality $n_{II}^- \geq n_I \geq n_{II}^+$. In this case, 100% spin polarized injection into region II will occur as long as the incident angle exceeds $\theta_{c1}$.

**Case II: The conduction band edge in region I is below that in region II ($V < 0$) and $m_{II} \geq m_I$ (electron's effective mass in the refraction medium is heavier)**

In this case, both $\theta_{c1}$ and $\theta_{c3}$ can have real values since both Equations (15) and (19) can be satisfied for certain electron energies.

In order to identify these energies, we start with Equation (15) which gives the condition for $\theta_{c1}$ to be real. Using the quantity $X = 1/\sqrt{E-V}$, as defined earlier, we can recast Equation (15) as

$$f_2 = -|V|X^2 - \frac{\sqrt{2m_I(\eta^2 + v^2)}}{\hbar}X + \left(1 - \frac{m_I}{m_{II}}\right) \leq 0 \quad (22)$$

The expression $f_2$ reaches its minimum value of $-\infty$ when $X = \infty$, i.e. when the electron energy $E = V$. At this energy (which corresponds to zero kinetic energy in region II), the expression $f_2$ is obviously a negative quantity (it is $-\infty$). Hence, Equation (22) is satisfied. However, when $E = \infty$, corresponding to $X = 0$, Equation (22) cannot be satisfied. Therefore, there are some electron energies that will make $\theta_{c1}$ real and some will not. If the electron energy is such that real values of $\theta_{c1}$ exist, then electron transmission into the corresponding spin eigenstate in the refraction medium will be blocked if the incident angle exceeds $\theta_{c1}$.

To examine if real values of $\theta_{c3}$ exist, we recast Equation (19) as

$$f_3 = -|V|X^2 + \frac{\sqrt{2m_I(\eta^2 + v^2)}}{\hbar}X + \left(1 - \frac{m_I}{m_{II}}\right) \leq 0 \quad (23)$$

The function $f_3$ assumes its minimum value at $X = \infty$ and its maximum value at $X = \frac{\sqrt{m_I(\eta^2 + v^2)}}{\sqrt{2}\hbar|V|}$. Equation (23) is obviously satisfied at the minimum value of the function. At the maximum value, the inequality in Equation (23) becomes the same as Equation (21) with $V$ replaced by $|V|$.

What we have proved for Case II is that for some electron energies, both critical angles $\theta_{c1}$ and $\theta_{c3}$ can exist (have real values). If the angle of incidence lies between the two critical angles, i.e. $\theta_{c3} > \theta_i > \theta_{c1}$, then 100% spin polarized injection into the refraction medium will take place.

We have not worked out the conditions for the case $m_{II} \leq m_I$, since the derivations are very similar. For this relationship between the effective masses, both $\theta_{c1}$ and $\theta_{c3}$ can have real values regardless of the sign of the potential step $V$. This is in contrast to the case when $m_{II} \geq m_I$ where $\theta_{c3}$ cannot have a real value if $V \geq 0$.

*2.1. Spin-polarized injection into the refraction medium*

We will consider only the case when $m_{II} \geq m_I$ and $V > 0$ (Case I) when $\theta_{c3}$ *does not exist*, but $\theta_{c1}$ may (at certain electron energies). If we can ensure that the electron energy is in an allowable range where $\theta_{c1}$ is real (hence exists) and the angle of incidence always exceeds $\theta_{c1}$, then an electron trying to transmit into the first spin eigenstate in the refraction medium will be totally reflected and the transmitted electrons will be exclusively of the other spin polarization, resulting in 100% spin-polarized injection. A possible construct to implement this is shown in Fig. 2. A 2-D trapezoidal mesa is etched in region I such that the angle ϕ subtended by its axis with the line demarcating the two regions exceeds $\theta_{c1}$. The incident energy of the electron can be tuned with a back gate (not shown) to make $\theta_{c1}$ real and less than or equal to ϕ. Injection (source) and detection (drain) electrodes are delineated at the locations shown, so that imposition of a chemical potential difference between the two electrodes will result in current flowing mostly parallel to the axis (i.e. the angle of incidence will be ϕ for most electrons, which exceeds the critical angle $\theta_{c1}$). Consequently, most electrons will transmit (refract) into only one spin state in region II.

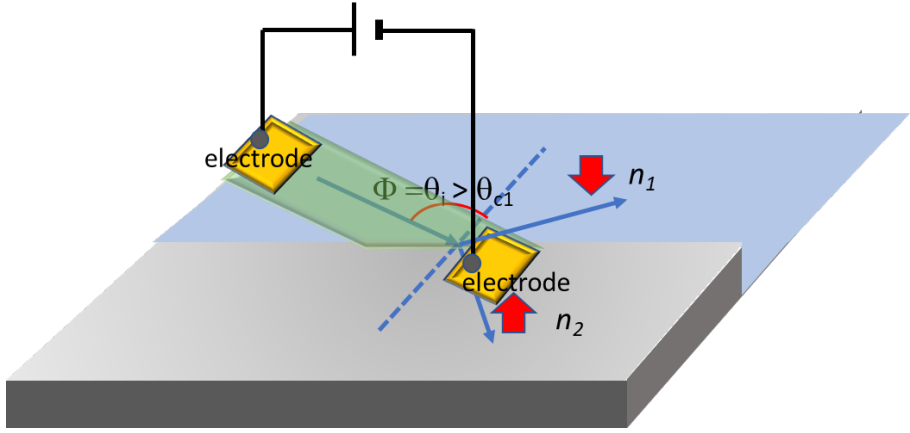

**Fig. 2:** Construct to cause spatial spin separation and spin polarization in the refraction medium.

As a related matter, we point out that there is another approach to spin polarized injection into region II, which is to apply a magnetic field in the *plane* of the 2-D system. In this case, the energy dispersion relations in region II become [10]

$$E_{\pm}^{II} = \frac{\hbar^2 \left(k_x^2 + k_z^2\right)}{2m_{II}} \\ \pm \sqrt{\left(g\mu_B B_z + \eta k_x - \nu k_z\right)^2 + \left(g\mu_B B_x + \eta k_z - \nu k_x\right)^2} \quad , \qquad (24)$$

where g is the Landé gyromagnetic ratio, $\mu_B$ is the Bohr magneton, and $B_x$, $B_z$ are the x- and z-components of the magnetic flux density. This dispersion relation is plotted in Fig. 3. If the transmitted (refracted) electron enters region II with energy E below the bottom of the higher spin-split subband, then the higher energy spin state will be evanescent while the lower energy spin state can propagate. This will also ensure spin-polarized injection into region II. A similar approach was advocated for a one-dimensional system earlier [12].

In this approach, spin polarized injection is accomplished by manipulating the incident electron energy and not by manipulating the angle of incidence. Because of the thermal spread in energy, we will require to satisfy the condition $g\mu_B B > kT$, where g is the Landau gyromagnetic ratio, $\mu_B$ is the Bohr magneton and B is the magnetic flux density. This would require a relatively large in-plane magnetic field which is not desirable in most cases and is best avoided. Moreover, one would require to reduce the spread in the transverse component of the wavevector $k_z$ (i.e. have a fixed $k_z$) which would still require a construct similar to Fig. 2.

Note that we have derived the laws of reflection and refraction from *energy conservation alone*, without invoking the continuity of the wave function and current across the interface. However, in order to obtain the transmission and reflection *amplitudes* into the two spin eigenstates, we will have to utilize the continuity conditions.

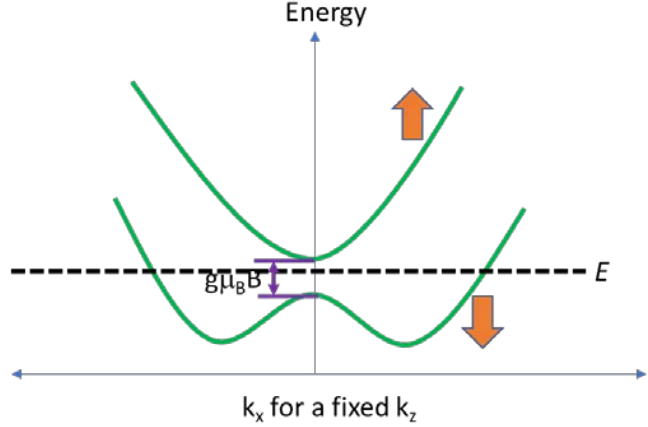

**Fig. 3:** Energy dispersion plots for the lowest spin-split subband in region II in the presence of an in-plane magnetic field.

## 3. Calculating the refraction and reflection amplitudes

The wave function of an electron in region I (with arbitrary spin polarization) can be written as

$$\Psi^I(x,z) = \begin{pmatrix} a \\ b \end{pmatrix} e^{i(k_x' x + k_z z)} + r \begin{pmatrix} a \\ b \end{pmatrix} e^{-i(k_x' x - k_z z)} + r' \begin{pmatrix} a' \\ b' \end{pmatrix} e^{-i(k_x' x - k_z z)}, \quad (25)$$

where the first term in the right hand side is the incident wave and the last two terms constitute the reflected wave, with $r$ being the reflection amplitude into the incident spin eigenstate and $r'$ the reflection amplitude into the orthogonal spin eigenstate $\left[ a^* a' + b^* b' = 0; \; |a|^2 + |b|^2 = |a'|^2 + |b'|^2 1 \right]$, where the asterisk denotes complex conjugate. There can be a spin-flip at the interface for the reflected electron (because of the spin-orbit interaction in the refraction medium) and hence we will have to consider that possibility here. The angle of reflection is always equal to the angle of incidence in region I regardless of whether or not there is a spin flip because $k_z' = k_z$ is a good quantum number.

The wave function of the refracted electron can be written as [11]

$$\Psi^{II}(x,z) = t_+ \begin{pmatrix} -\sin\varphi_{k+} \\ \cos\varphi_{k+} \end{pmatrix} e^{i(k_x^+ x + k_z z)} + t_- \begin{pmatrix} \cos\varphi_{k-} \\ \sin\varphi_{k-} \end{pmatrix} e^{i(k_x^- x + k_z z)} \quad (26)$$

where $t_+$ is the transmission amplitude into the first spin eigenstate and $t_-$ is that into the second spin eigenstate in the refraction medium. The *x*-components of the wave vectors in the two spin eigenstates are $k_x^\pm$. The angle $\varphi_{k\pm} = \frac{1}{2}\arctan\left(\frac{\nu k_x^\pm - \eta k_z}{\eta k_x^\pm - \nu k_z}\right)$. Note that the two eigenspinors in the two eigenstates are *not mutually orthogonal* since $\varphi_{k+} \neq \varphi_{k-}$. This happens because neither spin band has a fixed quantization axis in a 2-D system; the spin orientation varies with the wave vectors in both bands. At a *fixed wave vector*, the eigenspinors in the two bands will be mutually orthogonal, but at a *fixed energy* (which corresponds to different wave vectors in the two spin bands), the eigenspinors are not orthogonal. Note also that when only Rashba interaction is present, $\varphi_{k\pm} = -(1/2)\theta_r^{\pm}$, and when only Dresselhaus interaction is present, $\varphi_{k\pm} = \pi/4 - (1/2)\theta_r^{\pm}$.

The wave vectors in region II for an electron of energy $E$ incident from region I at an angle $\theta_i$ are given by the relations

$$k_x^+ = \frac{n_{II}^+}{n_I} n_I k_0 \sqrt{1 - \frac{n_I^2}{\left(n_{II}^+\right)^2} \sin^2 \theta_i}$$

$$= \frac{n_{II}^+}{n_I} \cdot \frac{\sqrt{2m_I(E-V)}}{\hbar} \sqrt{1 - \frac{n_I^2}{\left(n_{II}^+\right)^2} \sin^2 \theta_i} \qquad (27)$$

$$k_x^- = \frac{n_{II}^-}{n_I} n_I k_0 \sqrt{1 - \frac{n_I^2}{\left(n_{II}^-\right)^2} \sin^2 \theta_i}$$

$$= \frac{n_{II}^-}{n_I} \cdot \frac{\sqrt{2m_I(E-V)}}{\hbar} \sqrt{1 - \frac{n_I^2}{\left(n_{II}^-\right)^2} \sin^2 \theta_i}$$

and

$$k_x' = n_I k_0 \sqrt{1 - \sin^2 \theta_i} = \frac{\sqrt{2m_I(E-V)}}{\hbar} \sqrt{1 - \sin^2 \theta_i} \qquad (28)$$

$$k_z = n_I k_0 \sin \theta_i = \frac{\sqrt{2m_I(E-V)}}{\hbar} \sin \theta_i$$

The wave vector $k_x^+$ becomes imaginary when $\theta_i > \theta_{c1}$, making the angle $\varphi_{k+}$ complex, while the wave vector $k_x^-$ becomes imaginary when $\theta_i > \theta_{c3}$, making the angle $\varphi_{k-}$ complex. This feature represents the well-known fact that the refracted wave will be evanescent when the angle of incidence exceeds the critical angle.

Enforcing the continuity of the wavefunction at the junction between the two regions (at $x = 0$), we get that

$$t_+ \begin{pmatrix} -\sin \varphi_{k+} \\ \cos \varphi_{k+} \end{pmatrix} + t_- \begin{pmatrix} \cos \varphi_{k-} \\ \sin \varphi_{k-} \end{pmatrix} = (1+r) \begin{pmatrix} a \\ b \end{pmatrix} + r' \begin{pmatrix} a' \\ b' \end{pmatrix}, \qquad (29)$$

which can be re-written as

$$\begin{bmatrix} -\sin \varphi_{k+} & \cos \varphi_{k-} \\ \cos \varphi_{k+} & \sin \varphi_{k-} \end{bmatrix} \begin{bmatrix} t_+ \\ t_- \end{bmatrix} - \begin{bmatrix} a & a' \\ b & b' \end{bmatrix} \begin{bmatrix} r \\ r' \end{bmatrix} = \begin{bmatrix} a \\ b \end{bmatrix}. \qquad (30)$$

Enforcing the continuity of the spinor's derivative across the interface [6], we get that at $x = 0$

$$\left[ \frac{p_x}{m_{II}} - (\eta/\hbar)\sigma_z + (\nu/\hbar)\sigma_x \right] \Psi^{II}(x,z) \bigg|_{x=0} = \frac{p_{x'}}{m_I} \Psi^I(x,z) \bigg|_{x=0} \qquad (31)$$

which can be re-written as

$$\frac{\hbar k_x^+}{m_{II}} t_+ \begin{pmatrix} -\sin\varphi_{k+} \\ \cos\varphi_{k+} \end{pmatrix} + \frac{\hbar k_x^-}{m_{II}} t_- \begin{pmatrix} \cos\varphi_{k-} \\ \sin\varphi_{k-} \end{pmatrix} + \frac{\eta}{\hbar} t_+ \begin{pmatrix} \sin\varphi_{k+} \\ \cos\varphi_{k+} \end{pmatrix}$$
$$-\frac{\eta}{\hbar} t_- \begin{pmatrix} \cos\varphi_{k-} \\ -\sin\varphi_{k-} \end{pmatrix} + \frac{v}{\hbar} t_+ \begin{pmatrix} \cos\varphi_{k+} \\ -\sin\varphi_{k+} \end{pmatrix} + \frac{v}{\hbar} t_- \begin{pmatrix} \sin\varphi_{k-} \\ \cos\varphi_{k-} \end{pmatrix} \quad (32)$$
$$= [1-r]\frac{\hbar k_x'}{m_I}\begin{pmatrix} a \\ b \end{pmatrix} - r'\frac{\hbar k_x'}{m_I}\begin{pmatrix} a' \\ b' \end{pmatrix}$$

which can again be recast as

$$\begin{bmatrix} \alpha & \beta \\ \chi & \delta \end{bmatrix}\begin{bmatrix} t_+ \\ t_- \end{bmatrix} + \left(\hbar k_x'/m_I\right)\begin{bmatrix} a & a' \\ b & b' \end{bmatrix}\begin{bmatrix} r \\ r' \end{bmatrix} = \left(\hbar k_x'/m_I\right)\begin{bmatrix} a \\ b \end{bmatrix}$$

$$\alpha = -\left(\hbar k_x^+/m_{II} - \eta/\hbar\right)\sin\varphi_{k+} + (v/\hbar)\cos\varphi_{k+}$$
$$\beta = \left(\hbar k_x^-/m_{II} - \eta/\hbar\right)\cos\varphi_{k-} + (v/\hbar)\sin\varphi_{k-} \quad (33)$$
$$\chi = \left(\hbar k_x^+/m_{II} + \eta/\hbar\right)\cos\varphi_{k+} - (v/\hbar)\sin\varphi_{k+}$$
$$\delta = \left(\hbar k_x^-/m_{II} + \eta/\hbar\right)\sin\varphi_{k-} + (v/\hbar)\cos\varphi_{k-}$$

Defining new matrices as

$$[\mathbf{A}] = \begin{bmatrix} -\sin\varphi_{k+} & \cos\varphi_{k-} \\ \cos\varphi_{k+} & \sin\varphi_{k-} \end{bmatrix};\quad [\mathbf{B}] = \begin{bmatrix} a & a' \\ b & b' \end{bmatrix};\quad [\mathbf{C}] = \begin{bmatrix} a \\ b \end{bmatrix}$$

and $[\mathbf{A}'] = \begin{bmatrix} \alpha & \beta \\ \chi & \delta \end{bmatrix}$, we can re-write Equation (30) as

$$[\mathbf{A}]\begin{bmatrix} t_+ \\ t_- \end{bmatrix} - [\mathbf{B}]\begin{bmatrix} r \\ r' \end{bmatrix} = [\mathbf{C}], \quad (34)$$

and Equation (33) as

$$[\mathbf{A}']\begin{bmatrix} t_+ \\ t_- \end{bmatrix} + \frac{\hbar k_x'}{m_I}[\mathbf{B}]\begin{bmatrix} r \\ r' \end{bmatrix} = \frac{\hbar k_x'}{m_I}[\mathbf{C}]. \quad (35)$$

From Equation (34), we get

$$\begin{bmatrix} r \\ r' \end{bmatrix} = [\mathbf{B}]^{-1}[\mathbf{A}]\begin{bmatrix} t_+ \\ t_- \end{bmatrix} - [\mathbf{B}]^{-1}[\mathbf{C}]$$
$$= [\mathbf{B}]^\dagger[\mathbf{A}]\begin{bmatrix} t_+ \\ t_- \end{bmatrix} - [\mathbf{B}]^\dagger[\mathbf{C}] \quad (36)$$
$$= [\mathbf{B}]^\dagger[\mathbf{A}]\begin{bmatrix} t_+ \\ t_- \end{bmatrix} - \begin{bmatrix} 1 \\ 0 \end{bmatrix},$$

where the dagger denotes Hermitian conjugate. Note that the matrix $[\mathbf{B}]$ is unitary and hence its inverse is its Hermitian conjugate matrix.

Substituting the last result in Equation (35), we get

$$([\mathbf{A'}] + (\hbar k_x'/m_1)[\mathbf{A}]) \begin{bmatrix} t_+ \\ t_- \end{bmatrix} = 2(\hbar k_x'/m_1)[\mathbf{C}]$$

$$\Rightarrow \begin{bmatrix} t_+ \\ t_- \end{bmatrix} = 2(\hbar k_x'/m_1)([\mathbf{A'}] + (\hbar k_x'/m_1)[\mathbf{A}])^{-1}[\mathbf{C}] \quad (37)$$

$$\Rightarrow \begin{bmatrix} r \\ r' \end{bmatrix} = 2(\hbar k_x'/m_1)[\mathbf{B}]^\dagger[\mathbf{A}]([\mathbf{A'}] + (\hbar k_x'/m_1)[\mathbf{A}])^{-1}[\mathbf{C}] - \begin{bmatrix} 1 \\ 0 \end{bmatrix}.$$

Let us now define a new $2\times 2$ matrix $[\mathbf{D}] = 2(\hbar k_x'/m_1)([\mathbf{A'}] + (\hbar k_x'/m_1)[\mathbf{A}])^{-1} = \begin{bmatrix} d_{11} & d_{12} \\ d_{21} & d_{22} \end{bmatrix}$. Then from Equation (37), we get the transmission amplitudes in the three different intervals $[0^0, \theta_{c1}], [\theta_{c1}, \theta_{c3}], [\theta_{c3}, 90^0]$ for the incident angle $\theta_i$ as

$$t_+ = \begin{cases} d_{11}a + d_{12}b & (\text{for } 0^0 \leq \theta_i < \theta_{c1}) \\ 0 & (\text{for } \theta_{c1} \leq \theta_i < \theta_{c3}) \\ 0 & (\text{for } \theta_{c3} \leq \theta_i \leq 90^0) \end{cases}$$

$$t_- = \begin{cases} d_{21}a + d_{22}b & (\text{for } 0^0 \leq \theta_i < \theta_{c1}) \\ d_{21}a + d_{22}b & (\text{for } \theta_{c1} \leq \theta_i < \theta_{c3}) \\ 0 & (\text{for } \theta_{c3} \leq \theta_i \leq 90^0) \end{cases} \quad (38)$$

while the reflection amplitudes are found from Equation (37).

$$r = \begin{cases} [-a^* \sin\varphi_{k+} + b^* \cos\varphi_{k+}](d_{11}a + d_{12}b) \\ + [a^* \cos\varphi_{k-} + b^* \sin\varphi_{k-}](d_{21}a + d_{22}b) - 1 & (\text{for } 0^0 \leq \theta_i < \theta_{c1}) \\ [a^* \cos\varphi_{k-} + b^* \sin\varphi_{k-}](d_{21}a + d_{22}b) - 1 & (\text{for } \theta_{c1} \leq \theta_i < \theta_{c3}) \\ -1 & (\text{for } \theta_{c3} \leq \theta_i \leq 90^0) \end{cases}$$

$$(39)$$

$$r' = \begin{cases} [-a^{*} \sin\varphi_{k+} + b^{*} \cos\varphi_{k+}](d_{11}a + d_{12}b) \\ + [a^{*} \cos\varphi_{k-} + b^{*} \sin\varphi_{k-}](d_{21}a + d_{22}b) & (\text{for } 0^0 \leq \theta_i < \theta_{c1}) \\ [a^* \cos\varphi_{k-} + b^* \sin\varphi_{k-}](d_{21}a + d_{22}b) & (\text{for } \theta_{c1} \leq \theta_i < \theta_{c3}) \\ 0 & (\text{for } \theta_{c3} \leq \theta_i \leq 90^0) \end{cases}$$

Here we assumed that both critical angles exist (which is the most general case).

It is interesting to note that when the incident spin is not refracted into the refraction medium at all and suffers complete total internal reflection (i.e. when $\theta_{c3} \leq \theta_i \leq 90^0$), there is no spin flip during reflection because $r' = 0$. In this case, the incident spin does not "sample" the medium with spin-orbit interaction and hence there is no spin flip.

Equations (38) and (39) yield the values of the transmission and reflection amplitudes $t_+$, $t_-$, $r$ and $r'$ as functions of the incident angle $\theta_i$ for fixed $E$ and $V$, spin-orbit interaction strengths, effective masses, and incident spin polarization $\begin{pmatrix} a \\ b \end{pmatrix}$.

**Case of normal incidence:** Let us examine the special case when only Rashba interaction is present. At normal incidence, $\varphi_{k+} = \varphi_{k-} = 0$

Hence the matrix

$$[\mathbf{A}] = \begin{bmatrix} 0 & 1 \\ 1 & 0 \end{bmatrix} \text{ and } [\mathbf{A'}] = \begin{bmatrix} 0 & \hbar k_x^-/m_{II} - \eta/\hbar \\ \hbar k_x^-/m_{II} + \eta/\hbar & 0 \end{bmatrix}. \tag{40}$$

Therefore,

$$[\mathbf{A'}] + \hbar k_x'/m_I [\mathbf{A}]$$
$$= \begin{bmatrix} 0 & \hbar k_x'/m_I + \hbar k_x^-/m_{II} - \eta/\hbar \\ \hbar k_x'/m_I + \hbar k_x^+/m_{II} + \eta/\hbar & 0 \end{bmatrix}$$

and its inverse is the matrix

$$\begin{bmatrix} 0 & \left(\hbar k_x'/m_I + \hbar k_x^+/m_{II} + \eta/\hbar\right)^{-1} \\ \left(\hbar k_x'/m_I + \hbar k_x^-/m_{II} - \eta/\hbar\right)^{-1} & 0 \end{bmatrix}$$

Using the above results in Equation (37), we get

$$t_+ = \frac{2\hbar k_x' b}{m_I \left(\hbar k_x'/m_I + \hbar k_x^+/m_{II} + \eta/\hbar\right)} \tag{41}$$

and

$$t_- = \frac{2\hbar k_x' b}{m_I \left(\hbar k_x'/m_I + \hbar k_x^-/m_{II} - \eta/\hbar\right)}. \tag{42}$$

At normal incidence, $k_z = 0$ and hence the energy dispersion relations in the presence of Rashba interaction alone becomes (see Equation (2))

$$E_\pm^{II} = \frac{\hbar^2 k_x^2}{2m_{II}} \pm \eta k_x.$$

Since $k_x^+$ and $k_x^-$ are wavevectors in the two spin-split bands at the same energy, we get

$$\frac{\hbar^2 k_x^{+2}}{2m_2} + \eta k_x^+ = \frac{\hbar^2 k_x^{-2}}{2m_2} - \eta k_x^-$$

$$\Rightarrow \frac{\hbar^2\left(k_x^{+2} - k_x^{-2}\right)}{2m_2} + \eta\left(k_x^+ + k_x^-\right) = 0 \quad (43)$$

$$\Rightarrow \frac{\hbar k_x^+}{m_2} + \frac{\eta}{\hbar} = \frac{\hbar k_x^-}{m_2} - \frac{\eta}{\hbar}$$

Using the last relation in the expressions for the transmission amplitudes in Equations (41) and (42), we get that at normal incidence, $t_+ = \lambda b$ and $t_- = \lambda a$ where

$$\lambda = \frac{2\hbar k_x'}{m_1\left(\hbar k_x'/m_1 + \hbar k_x^+/m_2 + \eta/\hbar\right)}$$
$$= \frac{2\hbar k_x'}{m_1\left(\hbar k_x'/m_1 + \hbar k_x^-/m_2 - \eta/\hbar\right)}.$$

Hence, *if* the incident spin is polarized such that $|a| = |b|$ (which is the case, for example, when the spin is *x*-polarized or *y*-polarized), then $|t_+| = |t_-|$ when $\theta_i = 0$, i.e. at normal incidence. On the other hand, if the incident electron is *z*-polarized when either *a* or *b* is zero, then either $|t_-|$ or $|t_+|$ will vanish at normal incidence and 100% spin polarized injection into the refraction medium will take place.

Using Equations (41) – (43) in Equation (36), we also get that at normal incidence

$$r = \lambda - 1; \quad r' = 0, \quad (44)$$

which indicates that the electron never flips spin during reflection if it is normally incident on the interface. These features are borne out in the numerical examples given below.

## 4. Numerical Examples

For numerical examples, we consider pairs of materials in regions I and II similar to those considered in effective mass superlattices where the conduction band discontinuity at the interface is almost zero, i.e. $V = 0$, but the effective masses in the two regions are considerably different [13]. This case is picked arbitrarily and by no means implies that the theory works only when there is no potential discontinuity at the interface. The theory presented here works just as well when $V \neq 0$, except in that case, there will be an additional localized (delta-function) Rashba spin-orbit interaction at the interface due to the delta-function electric field at the interface caused by the step potential discontinuity. As long as the potential step is relatively small, this effect can be neglected. In the ensuing numerical examples, $V = 0$ and hence the results are exact.

**Case 1** $\left(m_{II} > m_I; \ V = 0\right)$ An example of a pair of materials conforming to this case is $Al_{0.5}In_{0.5}As_{0.49}Sb_{0.51}$ and GaSb which have effective masses of $0.067m_0$ and $0.049m_0$, respectively, and almost no conduction band offset [13]. We pick GaSb as region I and the alloy as region II. Rashba spin-orbit interaction in GaSb is considerably smaller than that in the alloy because of the presence of InAs in the latter, which is known to have strong Rashba coupling strength. Therefore we can assume (approximately) that region I has no spin-orbit interaction, but region II does. We ignore Dresselhaus interaction in both materials and assume that the Rashba interaction strength is $10^{-11}$ eV-m in $Al_{0.5}In_{0.5}As_{0.49}Sb_{0.51}$.

In Fig. 4, the solid and dashed curves show a plot of the energy dependence of the ratios $n_{II}^+/n_I$ and $n_{II}^-/n_I$, respectively, as obtained from Equation (11).

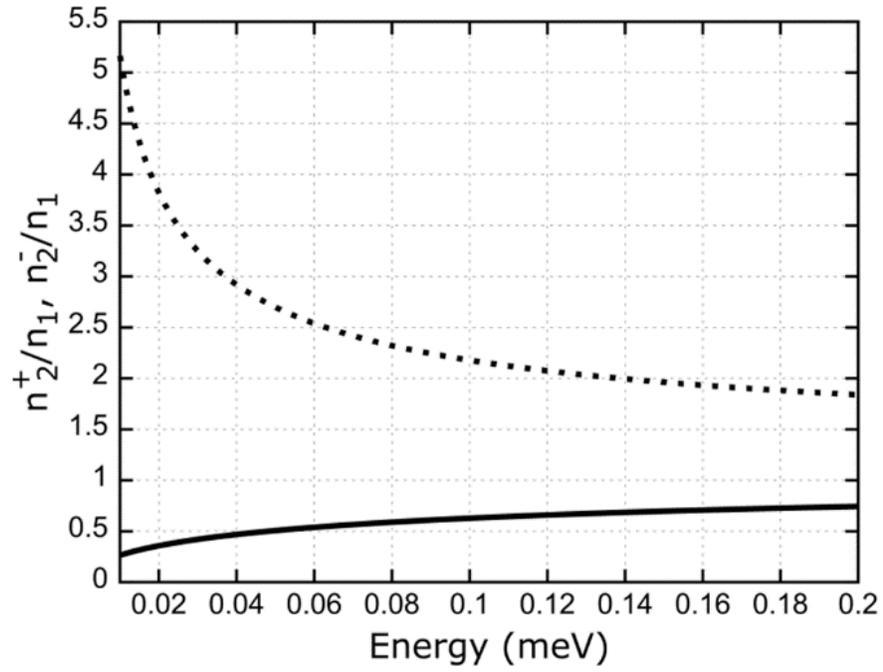

**Fig. 4:** Plot of $n_{II}^+/n_I$ (full line) and $n_{II}^-/n_I$ (dashed line) as a function of the energy of electron incident from region I.

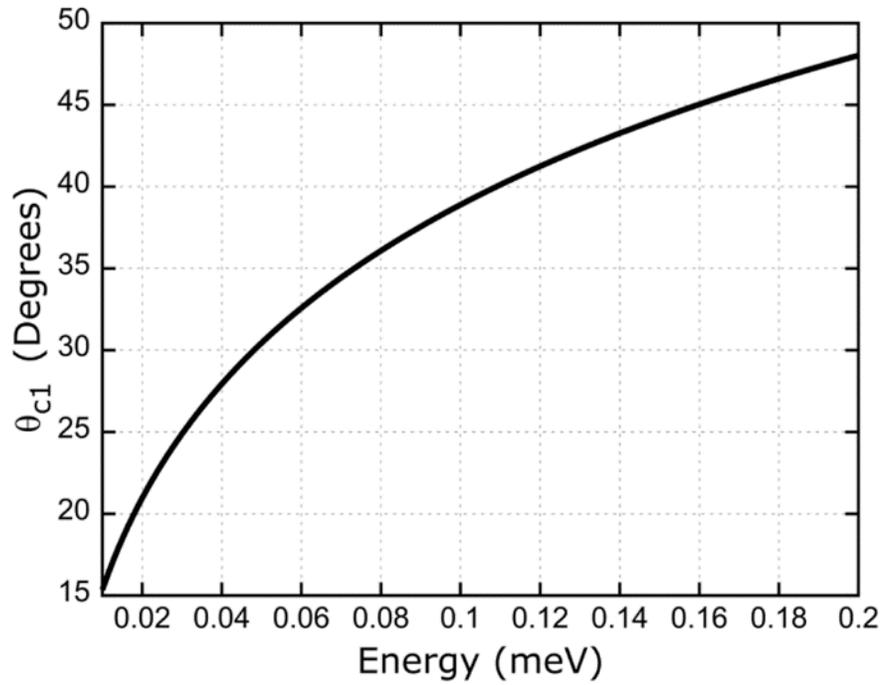

**Fig. 5:** Plot of the critical angle $\theta_{c1}$ as a function of the energy of the electron incident from region I. In this case, since the condition $n_{II}^- > n_I > n_{II}^+$ is satisfied for all incident energy, there is no real solution for $\theta_{c3}$.

We will assume that an electron is incident on the interface from region I with a Fermi energy $E_F = 0.05$ meV. The corresponding Fermi wavevector is $0.937 \times 10^7$ m$^{-1}$ in the alloy and the corresponding sheet carrier concentration is $1.4 \times 10^{13}$ m$^{-2}$. This is at least an order of magnitude higher than the intrinsic sheet carrier concentration of $10^{12}$ m$^{-2}$ in GaSb [14] and hence can be obtained with doping. From Equation (11), we get that for this system, $n_{II}^+/n_I = 0.51$ and $n_{II}^-/n_I = 2.73$ at the Fermi energy, which then satisfies the condition $n_{II}^- > n_I > n_{II}^+$. In this case, there is no real solution for $\theta_{c3}$, while $\theta_{c1} = 30.4^0$. A plot of the energy dependence of $\theta_{c1}$ is shown in Fig. 5.

Figures 6 and 7 show plots of the square magnitudes $|t_+|^2$, $|t_-|^2$, $|r|^2$, and $|r'|^2$ for two different spin polarizations of the incident electron: y-polarized (i.e., $a = 1/\sqrt{2}, b = 1/\sqrt{2}, a' = i/\sqrt{2}, b' = -i/\sqrt{2}$) and z-polarized (i.e., $a = 1, b = 0, a' = 0, b' = 1$) for the case discussed previously. As predicted by Equations (40) and (41), $|t_+|^2 = |t_-|^2$ when $\theta_i = 0$ (normal incidence) if the incident electron spin is y-polarized. If the incident electron is z-polarized, then $|t_+|^2 = 0$ when $\theta_i = 0$ (normal incidence). Furthermore, $|t_+|^2$ is exactly equal to zero (no refraction) when the incident angle is above $\theta_{c1}$ while $|t_-|^2$ remains non-zero, meaning that the refracted beam is completely spin polarized in that case. Above $\theta_{c1}$, a larger portion of the incident electron is reflected with the same polarization as the incident beam as $\theta_i$ increases. As shown in Fig. 6, $|t_-|^2$ reaches a maximum slightly above $\theta_{c1}$.

Note that $|t_+|^2$ and $|t_-|^2$ are *not* transmission probabilities and hence their values can exceed unity. They are simply related to the amplitudes of the refracted wave in region II and are defined through Equation (28). The actual transmission probabilities, defined as the ratio of the transmitted to incident current densities, will involve not just $|t_+|^2$ and $|t_-|^2$ but also the electron fluxes which are functions of the electron velocities. The expressions for the spin-dependent electron velocities are complicated in the presence of spin-orbit interaction and hence this discussion is omitted here (since it is outside the scope of this work).

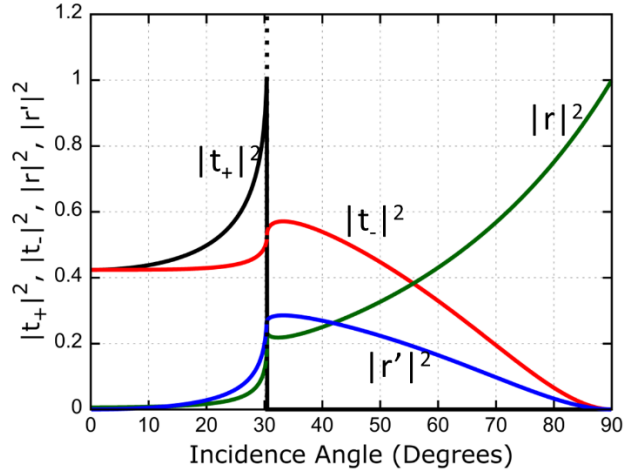

**Fig. 6:** Plots of the square magnitudes of the transmission and reflection $|t_+|^2, |t_-|^2, |r|^2$, and $|r'|^2$ as a function of the incident angle of the electron in region I. The energy of the incident electron is assumed to be $E = 0.05$ meV and the potential step at the interface $V = 0$. The effective mass in regions I and II are 0.049 $m_0$ and 0.067 $m_0$, respectively. The strength of the Rashba interaction η in region II is assumed to be $10^{-11}$ eV-m while Dresselhaus interaction is absent. The incident electron is assumed to be +y-polarized (i.e., $a = 1/\sqrt{2}, b = 1/\sqrt{2}, a' = i/\sqrt{2}, b' = -i/\sqrt{2}$). The black, red, green, and blue curve is a plot of the square magnitude $|t_+|^2, |t_-|^2, |r|^2$, and $|r'|^2$, respectively (color online).

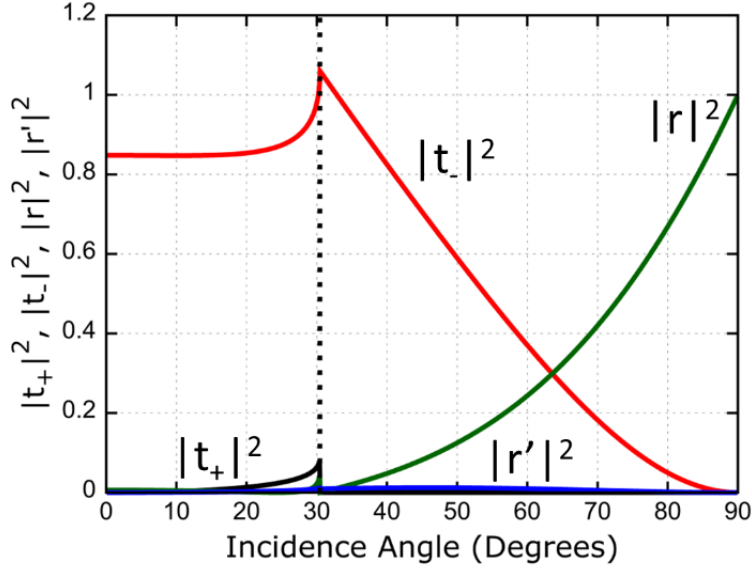

**Fig. 7:** Plots of the same quantities as in Fig. 6, except the incident electron is assumed to be +z-polarized (i.e., $a=1, b=0, a'=0, b'=1$). The black, red, green, and blue curve is a plot of the square magnitude $|t_+|^2, |t_-|^2, |r|^2$, and $|r'|^2$, respectively (color online).

Similar trends are seen when the incident beam is spin-polarized in the z-direction. In this case, $|t_-|^2$ is much larger right above $\theta_{c1}$ than in the case of a y-polarized incident beam. This would be the preferred mode of operation to achieve high degree of spin polarized current in region II.

**Case 2** $(m_{II} < m_I; V = 0)$: For this case, we could consider the following pair of materials, Ga$_{0.36}$In$_{0.14}$As$_{0.49}$Sb$_{0.22}$ ($m_I = 0.4 m_0$) and Al$_{0.23}$Ga$_{0.3}$In$_{0.47}$P ($m_{II} = 0.13 m_0$) with almost no conduction band offset ($V = 0$) [13]. Although in this case both regions can have comparable spin-orbit interaction strengths, we will ignore the Rashba interaction in the former and assume that in the latter, it is $10^{-11}$ eV-m. We will also ignore Dresselhaus interaction. These materials are chosen *for illustrative purposes only* and hence the assumptions regarding the relative strengths of spin-orbit interactions do not have to be accurate.

In Fig. 8, the solid and dashed curves show a plot of the energy dependence of the ratios $n_{II}^+/n_I$ and $n_{II}^-/n_I$, respectively, as obtained from Eq.(11). The difference between the two curves decreases as the energy of the incident electron increases.

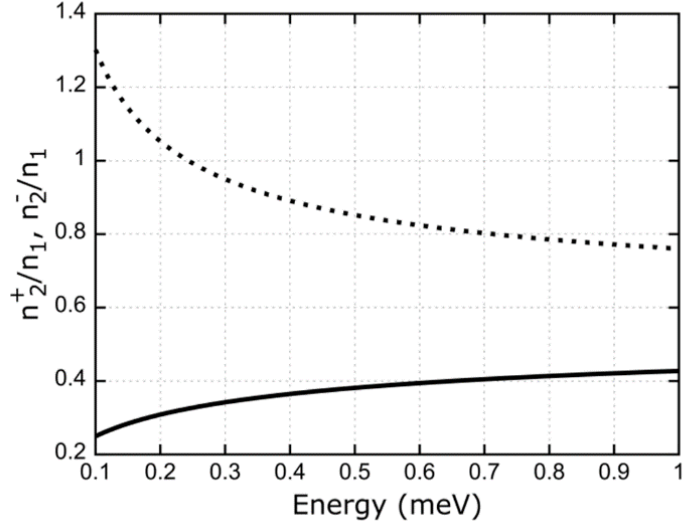

**Fig. 8:** Plot of $n_{II}^{+}/n_I$ (full line) and $n_{II}^{-}/n_I$ (dashed line) as a function of the incident electron energy $E$ for two materials with effective masses $m_I = 0.4 m_0$ and $m_{II} = 0.13 m_0$ with no potential step between them.

Assuming an electron incident from region I with a Fermi energy $E_F = 0.4$ meV, the corresponding Fermi wavevector is $6.5 \times 10^7$ m$^{-1}$ and the corresponding sheet carrier concentration is $6.7 \times 10^{14}$ m$^{-2}$. From Equation (11), we get that for this system, $n_{II}^{+}/n_I = 0.365$ and $n_{II}^{-}/n_I = 0.891$ at the Fermi energy. In this case, there are real solutions for *both* $\theta_{c1}$ and $\theta_{c3}$, with $\theta_{c1} = 21.4^0$ and $\theta_{c3} = 63^0$. A plot of the energy dependence of $\theta_{c1}$ and $\theta_{c3}$ is shown in Fig. 9.

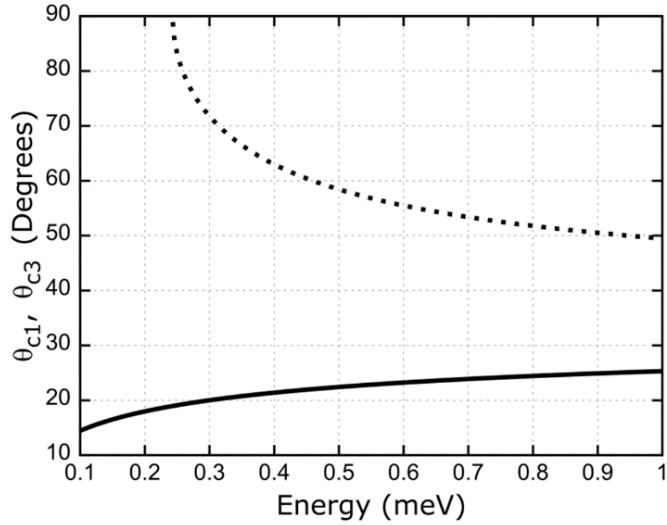

**Fig. 9:** Plots of the critical angles $\theta_{c1}$ (solid line) and $\theta_{c3}$ (broken line) as functions of the incident electron energy $E$. In this case, there are real solutions for $\theta_{c3}$ above a threshold incident energy of 0.243 meV.

Figures 10 and 11 show plots of the square magnitudes $|t_+|^2$, $|t_-|^2$, $|r|^2$, and $|r'|^2$ for two different spin polarizations of the incident electron: y-polarized (i.e. $a=1/\sqrt{2}, b=1/\sqrt{2}, a'=i/\sqrt{2}, b'=-i/\sqrt{2}$ in Fig. 10) and z-polarized (i.e., $a=1, b=0, a'=0, b'=1$, in Fig. 11) for an electron with incident energy $E = 0.4$ meV. As predicted by Equations (40) and (41), $|t_+|^2 = |t_-|^2$ when $\theta_i = 0$ (normal incidence) if the incident electron spin is y-polarized. If the incident electron is z-polarized, then $|t_+|^2 = 0$ when $\theta_i = 0$ (normal incidence). Furthermore, $|t_+|^2$ is exactly equal to zero when the incident angle is above $\theta_{c1}$ while $|t_-|^2$ remains non-zero, meaning that the refracted beam is completely spin polarized in that case. The coefficient $|r|^2$ eventually reaches unity above $\theta_{c3}$, i.e., the incident electron is totally reflected without flip of its spin polarization.

The numerical examples show some interesting features. Comparing Figures 6 and 7, or Figures 10 and 11, we immediately see that transmission into one spin eigenstate $|t_+|$ is much weaker than transmission into the other $|t_-|$ when the incident spin is z-polarized as opposed to y-polarized. One can understand this feature easily from Equations (41) and (42) which pertain to the special case of normal incidence. For +z-polarized spin, the incident spin $\begin{bmatrix}a\\b\end{bmatrix}=\begin{bmatrix}1\\0\end{bmatrix}$ and hence from Equations (41) and (42), $|t_+| = 0$ while $|t_-| > 0$. The reverse would have been true for –z-polarized spin for which $a = 0$ and $b = 1$. This trend is seen at all incident angles. Therefore, spin selective transmission into the refraction medium is more effective for z-polarized spin when averaged over all incident angles.

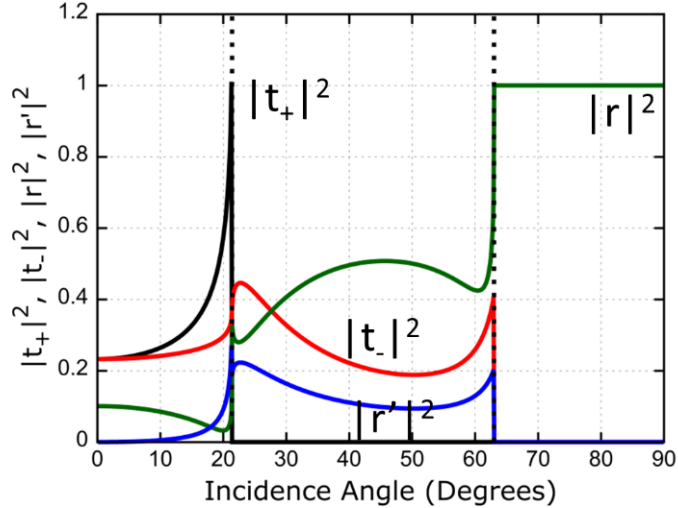

**Fig. 10:** Plots of the square magnitudes of the transmission and reflection $|t_+|^2, |t_-|^2, |r|^2$, and $|r'|^2$ as a function of the incident angle of the electron in region I. The energy of the incident electron is assumed to be $E= 0.4$ meV and the potential step at the interface $V = 0$. The effective mass in regions I and II are $m_I = 0.4 m_0$ and $m_{II} = 0.13 m_0$, respectively. The strength of the Rashba interaction $\eta$ in region II is assumed to be $10^{-11}$ eV-m. The incident electron is assumed to be +y-polarized (i.e., $a=1/\sqrt{2}, b=1/\sqrt{2}, a'=i/\sqrt{2}, b'=-i/\sqrt{2}$). The black, red, green, and blue curve is a plot of the square magnitude $|t_+|^2, |t_-|^2, |r|^2$, and $|r'|^2$, respectively (color online).

Another very interesting feature is that the likelihood of reflection with spin flip is much smaller than that without spin flip for *z*-polarized spin compared to *y*-polarized spin, especially when there is only one critical angle. The probability of reflection with spin flip is always zero at normal incidence, as can be seen from Equation (44).

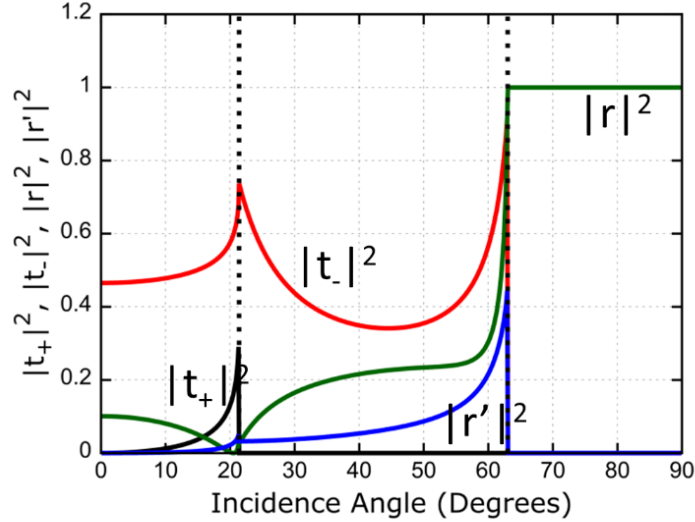

**Fig. 11:** Plots of the same quantities as in Fig. 10, except the incident electron is assumed to be +z-polarized (i.e., $a=1, b=0, a'=0, b'=1$). The black, red, green, and blue curve is a plot of the square magnitude $|t_+|^2, |t_-|^2, |r|^2$, and $|r'|^2$, respectively (color online).

## 5. Conclusion

In conclusion, we have derived the laws of reflection and refraction of a spin at the interface of two quasi-2-D regions – one with Rashba and Dresselhaus spin-orbit interaction and the other without – and found the conditions whereby refraction occurs only into one spin eigenstate of the refraction medium, causing spin polarized injection into the latter. These conditions are derived by invoking only energy conservation during the reflection/refraction process. The transmission (refraction) and reflection amplitudes are found from the continuities of the wavefunction and current across the interface between the two media.

The physics discussed in this paper causes spatial separation of two (mutually antiparallel) spins much in the same way as in the Stern-Gerlach experiment. A mesoscopic version of the Stern-Gerlach type filter by non-uniform spin-orbit interaction has been reported before [15], but generating non-uniform spin-orbit interaction is challenging. The experimental challenge here is less daunting. Another report proposes to leverage non-dispersive phases (Rashba and Aharonov-Bohm) to implement a Stern-Gerlach type device [16], but maintaining phase coherence is challenging at room temperature. Yet another work proposed using a T-shaped structure with spin-orbit interaction to spatially separate spins [17], which results in separation at specific kinetic energies of the incident electron. This construct requires ballistic transport, which is not a requirement in our case.

Note that the problem we explored in this paper is not a transport problem; it is an "interface" problem. The sharp interface is of zero physical extent and hence no transport can occur through the interface. Only what happens at the interface matters. What happens to the spin before reaching the interface (i.e. whether it suffers scattering, etc.) and what happens to the spin after it passes through the interface, are of no consequence and do not affect the laws of reflection and refraction. In the case of optics or electromagnetics, the laws of reflection and refraction are determined by the continuity of the electric and magnetic field components *at the interface only*, and what scattering the electromagnetic wave or light wave experiences before reaching the interface or after passing through the interface, does not affect Snell's law. The same is true here.

Finally, we conclude by stating that there is an emerging body of work dealing with spin injection in van der Waal's heterostructures, 2-D materials like graphene and $MoS_2$, and topological insulators like $Bi_2Se_3$. The Hamiltonian for the surface states of a topological insulator thin film looks like the Hamiltonian in Equation (1) without the Dresselhaus term, if we ignore higher order non-linear terms in the wavevector due to warping effects. With the warping effects included, these materials have different energy dispersion relations and different Hamiltonians than the ones considered here for a semiconductor quasi-2D layer. Therefore, treatment of these entities is outside the scope of the present work, but will be addressed by us in the future.